\documentclass[aps,prl,twocolumn,superscriptaddress,showpacs]{revtex4-1}
\usepackage{graphicx} 
\usepackage{newtxtext,newtxmath} 
\usepackage{braket}
\usepackage{xcolor}
\usepackage{gensymb}
\usepackage{siunitx}
\usepackage{float}
\definecolor{darkblue}{rgb}{0.1,0.2,0.6} 
\definecolor{darkred}{rgb}{0.8,0.1,0.2}
\usepackage[colorlinks,citecolor=darkblue,linkcolor=darkred,urlcolor=darkblue]{hyperref} 

\newcommand\simtight{\mkern1.5mu{\sim}\mkern1.5mu}
\newcommand\lessthantight{\mkern1.5mu{<}\mkern1.5mu}
\newcommand\approxtight{\mkern1.5mu{\approx}\mkern1.5mu}

\begin{document}

\title{Correlated Insulating States in Twisted Double Bilayer Graphene}

\author{G. William Burg}
\affiliation{Microelectronics Research Center, Department of Electrical and Computer Engineering, The University of Texas at Austin, Austin, TX 78758, USA}
\author{Jihang Zhu}
\affiliation{Department of Physics, The University of Texas at Austin, Austin, TX 78712, USA}
\author{Takashi Taniguchi}
\author{Kenji Watanabe}
\affiliation{National Institute of Materials Science, 1-1 Namiki Tsukuba, Ibaraki 305-0044, Japan}
\author{Allan H. MacDonald}
\email{macdpc@physics.utexas.edu}
\affiliation{Department of Physics, The University of Texas at Austin, Austin, TX 78712, USA}
\author{Emanuel Tutuc}
\email{etutuc@mail.utexas.edu}
\affiliation{Microelectronics Research Center, Department of Electrical and Computer Engineering, The University of Texas at Austin, Austin, TX 78758, USA}
\date{\today}

\begin{abstract}
We present a combined experimental and theoretical study of twisted double bilayer graphene with twist angles between 1\degree{} and 1.35\degree{}. Consistent with moir\'e band structure calculations, we observe insulators at integer moir\'e band fillings one and three, but not two.  An applied transverse electric field separates the first moir\'e conduction band from neighbouring bands, and favors the appearance of correlated insulators at 1/4, 1/2, and 3/4 band filling. Insulating states at 1/4 and 3/4 band filling emerge only in a parallel magnetic field ($B_{||}$), whereas the resistivity at half band filling is weakly dependent on $B_{||}$. Our findings suggest that 
correlated insulators are favored when a moir\'e flat band is spectrally isolated, and 
are consistent with a mean-field picture in which insulating state are established by 
breaking both spin and valley symmetries at 1/4 and 3/4 band filling and 
valley polarization alone at 1/2 band filling.
\end{abstract}

\maketitle

When two graphene sheets are stacked and twisted to an angle near $1.1\degree{}$, hybridization of the lowest lying energy bands gives rise to moir\'e superlattice bands
with a very flat dispersion \cite{bistritzer_moire_2011, suarez_morell_flat_2010}, which greatly enhances the local density of states and can induce strong electron-electron interactions. Recent advances in fabrication techniques \cite{kim_van_2016} have enabled van der Waals heterostructures in which the relative layer orientation is controlled to sub-degree precision. By using the relative twist between layers as a new design parameter, many interesting phenomena have been observed, including moir\'e bands and Hofstadter's butterfly spectra in twisted bilayer graphene \cite{cao_superlattice-induced_2016, kim_tunable_2017} and, most notably, correlated insulators \cite{cao_correlated_2018} and superconductivity \cite{cao_unconventional_2018, yankowitz_tuning_2019, lu_superconductors_2019} in 'magic angle' twisted bilayer graphene (MATBG).
 
Bernal stacked bilayer graphene has parabolic bands touching at low energies, four-fold spin and valley degeneracy, and a band structure that is strongly modified by a transverse electric field \cite{mccann_landau-level_2006, zhang_direct_2009}. Twisted double bilayer graphene (TDBG) is an attractive platform to probe electron-electron interactions in flat bands because its band widths and band gaps can be controlled by electrostatic gating. We present an electrical transport study of TDBG at angles between 1$\degree$ and 1.35$\degree$. We observe resistivity maxima corresponding to the single particle band gaps of the moir\'e band structure, as well as correlated insulators at half filling of the first conduction band at finite transverse electric fields. Measurements as a function of in-plane magnetic fields show insulators developing at $1/4$ and $3/4$ moir\'e band (MB) filling factor, suggesting spin polarization at MB quarter filling. By comparing the dependence of experimental data on twist angle and transverse electric field with band structure calculations, we conclude that correlated insulators are most likely to appear when a moir\'e band is spectrally separated from neighboring bands, with the moir\'e band flatness playing a secondary role.

Figure 1(a) shows a schematic of the moir\'e pattern formed by two Bernal stacked bilayer graphene sheets stacked with a relative twist ($\theta$). The moir\'e pattern retains the hexagonal structure of the underlying layers, and is characterized by a wave-length $\lambda=(a/2)/\sin(\theta/2)$, where $a = \SI{2.46}{\angstrom}$ is the graphene lattice constant. In reciprocal space, the Brillouin zone of the superlattice forms across the displaced $K$-points of the two graphene bilayers [Fig 1(b)].

\begin{figure}
\includegraphics[scale=0.355]{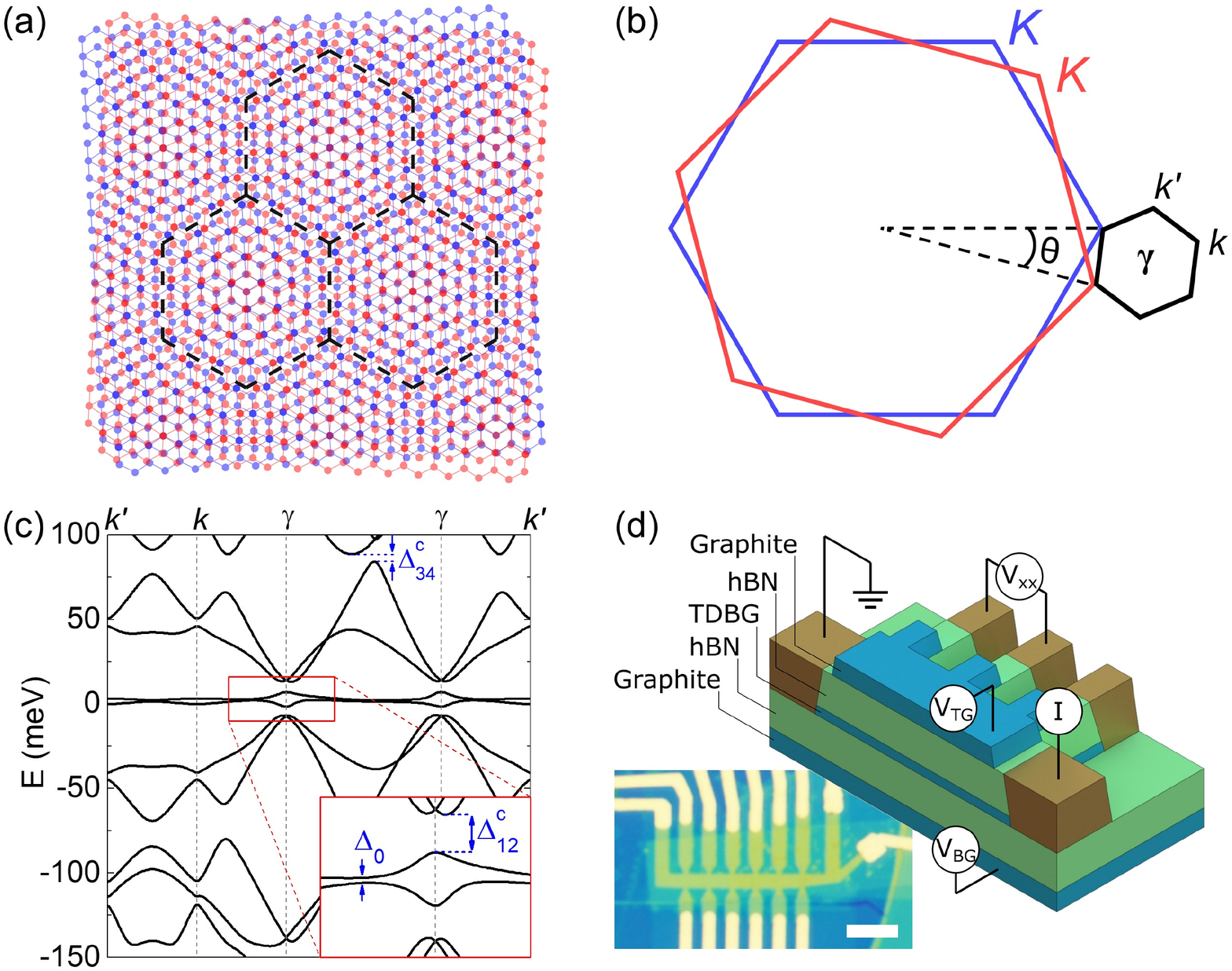}
\caption {\small{(a) Schematic of a moir\'e pattern formed by two Bernal stacked graphene bilayers twisted relative to one another. 
The dashed lines indicate the moir\'e lattice. (b) The moir\'e Brillouin zone of TDBG (black) near the corners of the two bilayer graphene Brillouin zones (red, blue). (c) TDBG band structure calculated for $\theta= 1.01\degree$. The inset shows a zoomed view of the lowest energy bands near the $\gamma$ point. Gaps at charge neutrality and between conduction sub-bands are labeled. Similar gaps are present between the corresponding valence sub-bands. (d) Schematic of the Hall bar structure used to probe TDBG samples, with top and bottom graphite gates. The inset shows an optical micrograph of a typical sample, with a 5 $\mu$m scale bar.}}
\label{Fig1}
\end{figure}

Figure 1(c) shows the moir\'e band structure calculated for TDBG with $\theta=1.01\degree$,
assuming that the middle two graphene layers are coupled by the same sub-lattice and position dependent hybridization as in twisted bilayer graphene \cite{bistritzer_moire_2011, Rafi_2010}. The $\mathbf{k}$-dependent Hamiltonian is then constructed in a plane-wave representation. To be consistent with atomistic calculations that account for 
strain and out-of-plane relaxation effects\cite{Taa_Tab_Wijk_2015, Taa_Tab_Jain_2016}, 
we set $|T_{AA}|/|T_{AB}|=0.8$, where $T_{AA}$ and $T_{AB}$ are respectively the interlayer hopping amplitudes between A and A and between A and B sublattices.
The outer two graphene layers are described by the minimal model of Bernal stacked bilayer graphene, considering only the hopping ($\gamma_1$) between dimer sites. 
We use $\gamma_1 = 0.33$ eV, consistent with infrared spectroscopy studies \cite{AB_bilayer_gamma1_2008, AB_bilayer_gamma1_2009}. For twists in the 1-1.3\degree{} range, the lowest lying conduction and valence bands become relatively flat, which promotes electron-electron interactions and can lead to correlated transport phenomena \cite{cao_correlated_2018, cao_unconventional_2018, yankowitz_tuning_2019, sharpe_emergent_2019, lu_superconductors_2019}. Figure 1(c) exhibits gaps at neutrality between the lowest conduction and valence bands ($\Delta_\mathrm{0}$), and also gaps within the conduction ($c$) and valence ($v$) band structures between the first and second ($\Delta_\mathrm{12}^{c,v}$), and third and fourth MBs ($\Delta_\mathrm{34}^{c,v}$).  The mechanism responsible for band flatness in TDBG is quite distinct from that in twisted single-layer systems, as explained more fully in the supplemental material.

Our TDBG samples are encapsulated in 20-50 nm thick boron nitride with graphite top and bottom gates [Fig 1(d)]. All layers are mechanically exfoliated, and the TDBG is realized using techniques similar to those described in \cite{kim_van_2016}, with both graphene bilayers originating from the same single crystal and subsequently twisted to a precise angle during the transfer process. The TDBG samples are shaped into Hall bars using reactive ion etching, and independent edge contacts \cite{wang_one-dimensional_2013} are made to the active area and gates. We discuss three TDBG samples, with twist angles $\theta =$ $1.01\degree{}$, $1.10\degree$, and $1.33\degree{}$. The $1.33\degree{}$ sample has a uniform twist angle to within $\lessthantight{}0.01\degree{}$ over a channel length of $\SI{5.5}{\mu{}m}$, while the $1.10\degree$ sample shows a variation in the twist angle of $\simtight{} 0.015\degree{}/\mu{}\mathrm{m}$ along the channel. The $1.01\degree{}$ sample did not have a sufficient number of contacts to quantify the angle uniformity.

\begin{figure}
\includegraphics[scale=0.385]{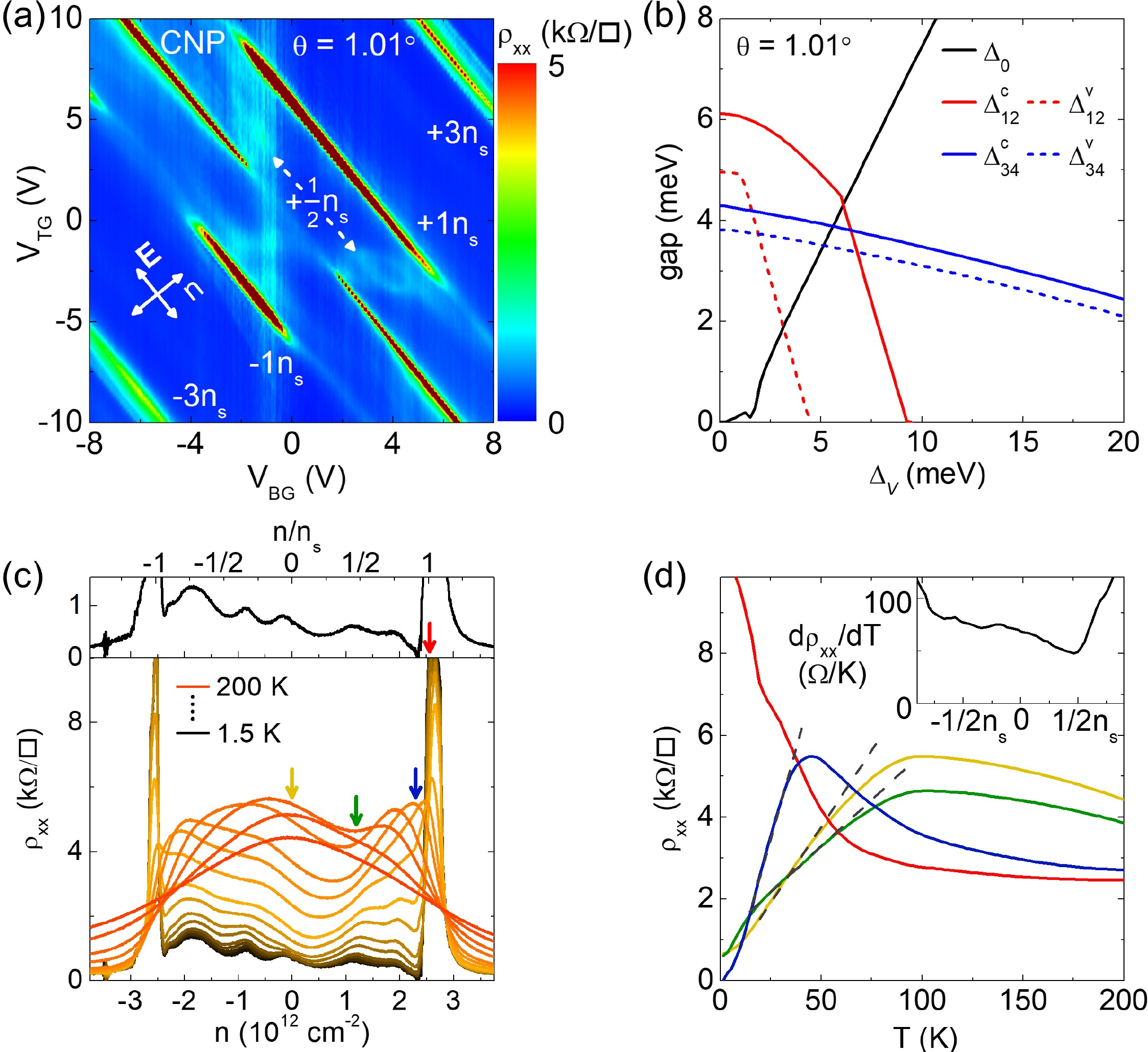}
\caption {\small{(a) Contour plot of $\rho_\mathrm{xx}$ vs. $V_\mathrm{TG}$ and $V_\mathrm{BG}$ measured at \SI{1.5}{K} in TDBG with $\theta = 1.01\degree$. Resistivity maxima are observed at charge neutrality, $\pm$$1n_\mathrm{s}$, and $\pm$$3n_\mathrm{s}$, as well as at +1/2$n_\mathrm{s}$. The inset axes indicate the directions of varying $E$ and $n$ values. (b) Calculated band gaps between different moir\'e conduction (solid) and valence (dashed) band as a function of the on-site energy difference between layers, $\Delta_V$. (c) $\rho_\mathrm{xx}$ vs. $n$
measured at different $T$ values, at $V_\mathrm{TG}$ = -0.8 V. Top inset: zoomed view of $\rho_\mathrm{xx}$ vs. $n_\mathrm{s}$ at 1.5 K, which shows several resistivity maxima at fractional band fillings. (d) $\rho_\mathrm{xx}$ vs. $T$ at different band fillings marked by arrows in panel (c). Dashed lines are a guide to the eye. Inset: $dR_\mathrm{xx}$/$dT$ vs. $n$ in units of $n_\mathrm{s}$ in the linear region.}}
\label{fig2}
\end{figure}

The longitudinal ($\rho_\mathrm{xx}$) and Hall ($\rho_\mathrm{xy}$) resistivities of the TDBG samples are probed using low frequency lock-in techniques.  Our dual gated structure allows top ($V_\mathrm{TG}$) and bottom ($V_\mathrm{BG}$) gate voltages to independently tune the carrier density, $n = (V_\mathrm{TG}C_\mathrm{TG} + V_\mathrm{BG}C_\mathrm{BG})/e$, and the transverse electric field, $E = (V_\mathrm{TG}C_\mathrm{TG} - V_\mathrm{BG}C_\mathrm{BG})/2\varepsilon_\mathrm{0}$, where $C_\mathrm{TG}$ and $C_\mathrm{BG}$ are the top and bottom gate capacitances, $e$ is the electron charge, and $\varepsilon_\mathrm{0}$ is the vacuum permittivity. Figure 2(a) shows $\rho_\mathrm{xx}$ vs $V_\mathrm{TG}$ and $V_\mathrm{BG}$ in the 1.01$\degree$ sample at a temperature $T = \SI{1.5}{K}$. The data show resistivity maxima at the charge neutrality point (CNP), and at a fixed density $n_\mathrm{s}$ and 3$n_\mathrm{s}$. Based on Fig. 1(c), which show gaps at neutrality as well as between the the first and second, and third and fourth MBs,  we associate $n_\mathrm{s}$ with filling of the first MB. Using the four-fold spin and valley degeneracy for each MB state, the wave-length, and therefore the twist angle, can be determined using $\lambda^2 = 4(2/\sqrt{3}n_\mathrm{s})$.

Interestingly, $\rho_\mathrm{xx}$ at the CNP increases as a function of $E$-field, while the $\rho_\mathrm{xx}$ values at $\pm1n_\mathrm{s}$ are large at $E=0$, but decrease with increasing $E$. The resistivity at $\pm3n_\mathrm{s}$ is relatively constant within the accessible gate voltage range. In addition, we observe resistivity peaks at fractional MB fillings, most notably at half filling of the first conduction band within two small $E$-field windows. 
The appearance of a gap in a half filled band is a hallmark of strong correlation physics, as demonstrated explicitly in ultra-cold atoms trapped in optical lattices \cite{greiner_quantum_2002, stoferle_transition_2004, jordens_mott_2008}.

To help elucidate the TDBG transport characteristics, we examine the band structure evolution as a function of $E$-field by introducing an on-site energy difference between adjacent layers ($\Delta_\mathrm{V}$), which should be viewed as the external 
field potential difference $\Delta_\mathrm{V}=eEd$ corrected for screening; $d$ is the graphene layer separation. The screening effect due to induced carriers is non-negligible (see supplemental material), and results in a significant reduction in $\Delta_\mathrm{V}$ compared to $eEd$.
Figure 2(b) shows how the gaps introduced in Fig. 1(c) vary with $\Delta_\mathrm{V}$ when $\theta=1.01\degree$. $\Delta_\mathrm{0}$ behaves like the gap of a single graphene bilayer and increases with $E$ \cite{zhang_direct_2009}.  On the other hand, $\Delta_\mathrm{12}^{c,v}$ decreases and vanishes at a finite $E$, while $\Delta_\mathrm{34}^{c,v}$ remains relatively constant. For both $\Delta_\mathrm{12}^{c,v}$ and $\Delta_\mathrm{34}^{c,v}$, the gaps between valence sub-bands are smaller and tend to zero at lower $E$-fields than for the corresponding conduction sub-bands. A comparison between Fig. 2(a) and 2(b) data shows good qualitative agreement between experiment and calculations, and validates the assignment of $n_\mathrm{s}$ to full MB filling, and therefore also the angle extraction from the transport characteristics. Interestingly, the $\rho_\mathrm{xx}$ maxima at $n_\mathrm{s}/2$ is observed at $E$-field values that yield insulators at both charge neutrality and $+1n_\mathrm{s}$, {\it i.e.} when $\Delta_\mathrm{0} \simeq \Delta_\mathrm{12}^{c}$. Under this condition, the first conduction band is maximally separated from both of its neighboring bands, suggesting that isolated bands favor the emergence of correlated insulators, and localized mainly in the outside layer that has the highest on-site energy (Fig. S1). This observation is also consistent with the absence of features at half filling in the valence band of Fig. 2(a) data, since $\Delta_\mathrm{0}$ and $\Delta_\mathrm{12}^{v}$ are never large enough at the same $E$-field to sufficiently separate the first valence band from surrounding bands.

In Figures 2(c-d) we examine the temperature dependence of the $\theta=1.01\degree{}$ TDBG. Figure 2(c) shows $\rho_\mathrm{xx}$ vs $n$ along a horizontal line-cut of Fig. 2(a) data, at $V_\mathrm{TG} = \SI{-0.8}{V}$, from $\SI{1.5}{K}$ to $\SI{200}{K}$. The upper inset shows a zoomed view of the line-cut at $T = \SI{1.5}{K}$ in which several developing $\rho_\mathrm{xx}$ maxima are observed at half and quarter MB fillings. For densities between the $\pm1n_\mathrm{s}$ insulators, the temperature dependence is metallic at lower $T$'s, with $\rho_\mathrm{xx}$ increasing with $T$ up to $\simtight{}\SI{50}{K}$, followed by a decrease with increasing $T$. Figure 2(d) shows the $T$-dependence at select densities, indicated by the arrows in Fig 2(c). In the metallic regions the resistivity has a nearly linear dependence on temperature, similar to experimental observations in MATBG \cite{cao_strange_2019, polshyn_phonon_2019}, which has been theoretically attributed either to acoustic phonon scattering in flat bands \cite{wu_phonon-induced_2019, li_phonon_2019} or to strong correlation effects \cite{cao_strange_2019}. 
The decrease in $\rho_\mathrm{xx}$ at higher temperatures is attributed to thermal activation to higher bands which are more dispersive and therefore have higher electron velocities \cite{polshyn_phonon_2019}. 
We note that exceptionally low $\rho_\mathrm{xx}$ values can be observed at certain MB fillings [blue arrow and trace in Fig. 2(c) and 2(d), respectively], which might signal an emerging superconducting state. Figure 2(d) inset shows that the slope of $\rho_\mathrm{xx}$ vs. $T$ in the linear regions is relatively agnostic to $n$, apart from slight dips indicating emergent insulators at fractional band fillings.   

\begin{figure}
\includegraphics[scale=0.37]{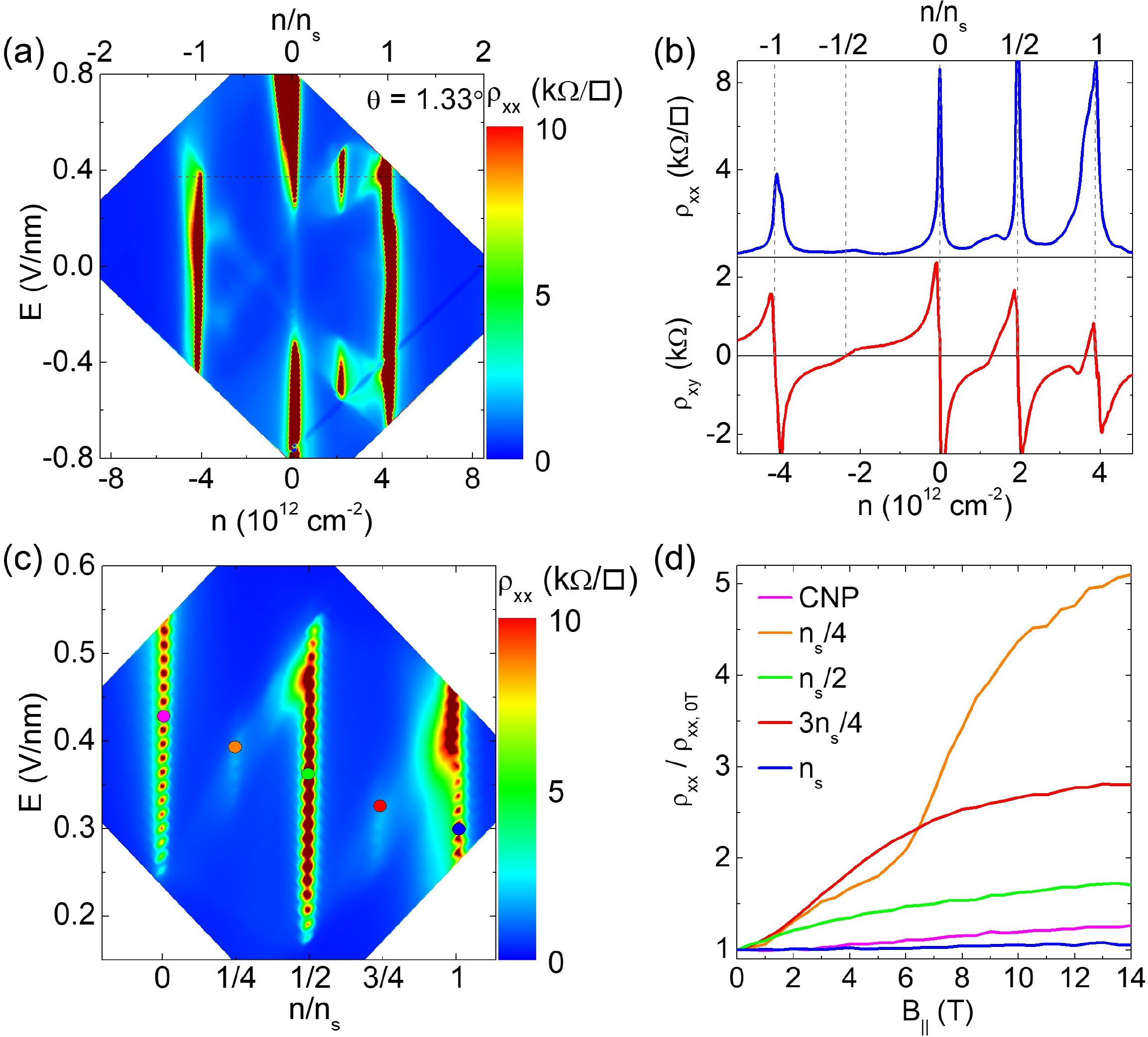}
\caption {\small{(a) $\rho_\mathrm{xx}$ vs. $E$ and $n$ measured in a TDBG sample with $\theta=1.33\degree$, at $T=\SI{1.5}{K}$. The top axis shows $n$ in units of $n_\mathrm{s}$. (b) $\rho_\mathrm{xx}$ and $\rho_\mathrm{xy}$ vs. $n$ measured at $B = \SI{1}{T}$, and $E \simeq \SI{0.4}{V/nm}$ [dashed line in panel (a)]. The dashed lines indicate $\rho_\mathrm{xx}$ maxima and $\rho_\mathrm{xy}$ sign changes at half or full MB fillings. (c) $\rho_\mathrm{xx}$ vs. $E$ and $n/n_\mathrm{s}$ measured at $B_\parallel = \SI{14}{T}$, and $T=1.5$ K. Insulating states develop at $n_\mathrm{s}/4$ and $3n_\mathrm{s}/4$. (d) $\rho_\mathrm{xx}$ normalized to the zero field value vs. $B_\mathrm{||}$ at different band fillings, as marked in panel (c).}}
\label{fig3}
\end{figure}

 Figure 3(a) shows $\rho_\mathrm{xx}$ vs $n$ and $E$ measured in a TDBG sample with $\theta=1.33\degree$, at $T=1.5$ K. Similar to the Fig. 2(a) data, single particle gaps appear at charge neutrality and $\pm1n_\mathrm{s}$, and $n_\mathrm{s}/2$ correlated insulators arise in a range of positive and negative $E$-fields. The $\rho_{xx}$ vs. $T$ data of the $\theta = 1.33\degree$ sample is similar to Fig. 2(d) data, showing a linear dependence below $\SI{50}{K}$, with density dependent $d\rho_{xx}/dT$ values of $\approxtight{}\SI{100}{\Omega/K}$ (Fig. S2). Figure 3(b) shows $\rho_\mathrm{xx}$ and $\rho_\mathrm{xy}$ vs. $n$ measured in a perpendicular magnetic field $B = \SI{1}{T}$ at $E \approx \SI{0.4}{V/nm}$ [dashed line in Fig 3(a)]. The $\rho_\mathrm{xy}$ vs. $n$ data changes sign at the single particle and correlated insulators, consistent with a transition between hole-like and electron-like bands when the Fermi level crosses an energy gap, and signaling that the four-fold band degeneracy is lifted in the first conduction MB.  As in the Fig. 2 data, the insulators at $n_\mathrm{s}/2$ are present at $E$-field values concomitant with insulating states at both CNP and $+n_\mathrm{s}$. However the correlated insulators at $n_\mathrm{s}/2$ are now more prominent, with $\rho_\mathrm{xx}$ values similar to those of the single particle insulators.

Band structure calculations offer an explanation for the differences between the $\theta=1.01\degree$ and $\theta=1.33\degree$ samples (Fig. S3). As a function of $E$-field, the $\Delta_\mathrm{0}=\Delta_\mathrm{12}^{c}$ condition provides the maximum separation of the first MB from the neighbouring bands. This separation varies with $\theta$, and reaches a maximum at $\theta\approx1.3-1.4\degree$.  We note that the width of the first conduction band at $\Delta_\mathrm{0} = \Delta_\mathrm{12}^{c}$ also increases with $\theta$.  Together the experimental data and calculations suggest that 
correlated insulators at fractional band fillings are most likely to emerge if the band is maximally separated from neighbouring bands, and that this criterion is more important than extreme band flatness. 
If only band flatness is 
considered, correlated insulators should be more prominent at $\theta\simeq 1\degree$ where the lowest bands are narrower.  We emphasize again that the mechanism for band flatness in TDBG is different than in the twisted single layer case, and that the band width in TDBG is a less sensitive function of twist angle.

\begin{figure*}
\includegraphics[scale=0.645]{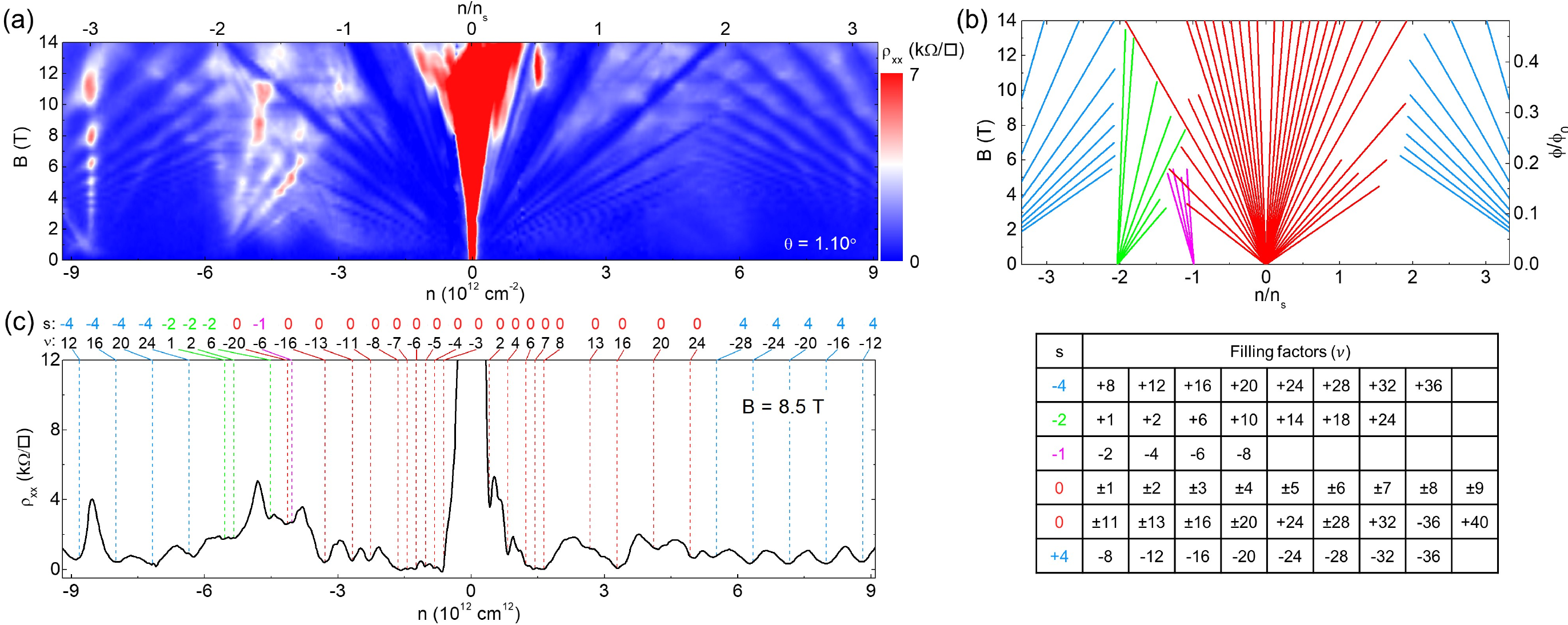}
\caption {\small{(a) $\rho_\mathrm{xx}$ vs $n$ and $B$ measured in a TDBG sample with $\theta = 1.10\degree$ at \SI{0.3}{K}, which shows the Hofstadter's butterfly with fans originating from the CNP, $-1n_\mathrm{s}$, $-2n_\mathrm{s}$, and $\pm4n_\mathrm{s}$, and emergent fans at -3$n_\mathrm{s}$, -3/2$n_\mathrm{s}$, and +2$n_\mathrm{s}$. (b) Summary of the fans observed in panel (a) data. The right $y$-axis is the magnetic flux per moir\'e unit cell ($\phi$) in units of the flux quantum ($\phi_0=h/e$). The lower table shows the filling factors for each fan. (c)  $\rho_\mathrm{xx}$ vs $n$ measured at $B = 8.5$ T. Dashed lines indicate the originating sub-band and Landau level filling factor for each identifiable minimum. Colors correspond to the fans in panel (b).}}
\label{fig4}
\end{figure*}

Further insight into the nature of the correlated insulators is provided by measurements in an in-plane magnetic field ($B_\mathrm{\parallel}$), which couples only to the electron spin.
Figure 3(c) shows $\rho_\mathrm{xx}$ vs $E$ and $n$ at $B_\mathrm{\parallel} = \SI{14}{T}$, and $T=1.5$ K. Correlated insulators emerge at $n_\mathrm{s}/4$ and $3n_\mathrm{s}/4$ in narrower $E$-field windows of $\approxtight{}\SI{0.05}{V/nm}$. In addition, the insulator at $n_\mathrm{s}/2$ extends over a slightly larger range of $E$-fields. The presence of gaps at each MB quarter filling suggests a full lifting of the four-fold spin and valley degeneracy, in which each electron added per moir\'e unit cell is polarized in both spin and valley.  Figure 3(d) shows $\rho_\mathrm{xx}$ normalized to its zero field value ($\rho_\mathrm{xx,0T}$) vs. $B_\mathrm{\parallel}$ at different MB fillings.
At $n_\mathrm{s}/4$ and $3n_\mathrm{s}/4$, $\rho_\mathrm{xx}$ grows rapidly with $B_\parallel$, whereas the 
growth is more gradual at $n_\mathrm{s}/2$. 
Figure 3(c-d) data suggest the insulators at $n_\mathrm{s}/4$ and $3n_\mathrm{s}/4$ are spin polarized, with gaps enhanced by the applied $B_\mathrm{\parallel}$, and correspondingly that the insulator at $n_\mathrm{s}/2$ is valley polarized.

We also consider the effect of a magnetic field perpendicular to the sample plane. Electrons in a periodic potential and perpendicular magnetic field develop a self-similar energy spectrum known as Hofstadter's butterfly \cite{hofstadter_energy_1976}, which has been studied extensively in graphene/boron nitride moir\'e patterns \cite{ponomarenko_cloning_2013, dean_hofstadter/s_2013, hunt_massive_2013, yu_hierarchy_2014}, and in twisted bilayer graphene \cite{cao_superlattice-induced_2016, kim_tunable_2017, cao_correlated_2018, cao_unconventional_2018, yankowitz_tuning_2019, lu_superconductors_2019}. In such systems, quantum Hall states (QHSs), indexed by a Landau level filling factor ($\nu$) and a sub-band filling factor ($s$), emerge when $n=\nu\cdot(eB/h) + s\cdot n_\mathrm{s}$; $e$ is the electron charge and $h$ is Planck's constant. Figure 4(a) shows a contour plot of $\rho_\mathrm{xx}$ vs. $n$ and $B$ measured in a TDBG sample with $\theta = 1.10\degree{}$, at $T = \SI{0.3}{K}$. The data shows QHS fans corresponding to $s = 0, -1, -2, \pm4$, with emerging fans at $s = -3, -3/2$, and $+2$. While the $\nu$ values are predominately multiples of four, the $s = 0$ fan shows $\rho_\mathrm{xx}$ minima at single integer filling factors, which indicates a lifting of the Landau level spin and valley degeneracy, as well as unexpected states at $\nu = \pm11$ and $\pm13$ [Fig. 4(b)]. In Fig. 4(c) we show a line-cut of Fig. 4(a) along $B = \SI{8.5}{T}$ where the various minima are labeled according to MB and $\nu$. At high fields, the $s=0$ fan exhibits $\rho_\mathrm{xx}$ minima corresponding to $\nu = \pm1/2$ (Fig. S4). It is also interesting to examine the fan intersections. For example, where the $s = 0$ and $s = -2$ fans meet, only the $\nu = -16$ minima from the $s = 0$ fan persists while all others are suppressed. A similar behavior is observed for $\nu = +16$. This may point towards topological distinctions between the different moir\'e bands or Landau levels \cite{lian_landau_2018}.

We have presented a study of electrical transport in twisted double bilayer graphene, a system that exhibits single particle gaps in the moire band spectrum along with correlated insulators at half and quarter fillings of the first conduction band. A combination of theoretical and experimental data suggests that correlated insulators are most likely to occur in a band when it is maximally separated from neighbouring bands. Measurements in an in-plane magnetic field indicate that the correlated insulators are predominantly spin polarized at 1/4 and 3/4 fillings, and valley polarized at 1/2 filling.

\medskip
\begin{acknowledgments}
This work was supported by the National Science Foundation grants EECS-1610008 and DMR-1720595, Army Research Office under Award W911NF-17-1-0312, and the Welch Foundation. Work was partly done at the Texas Nanofabrication Facility supported by NSF grant NNCI-1542159. K.W. and T.T. acknowledge support from the Elemental Strategy Initiative conducted by the MEXT, Japan and JSPS KAKENHI Grant Numbers JP15K21722.
\end{acknowledgments}

\noindent
\textit{Note added.}---During the preparation of this manuscript we became aware of two related studies \cite{liu_spin-polarized_2019, cao_electric_2019}.


%

\clearpage
\onecolumngrid
\section{Supplemental Material:\\ Correlated Insulating States in Twisted Double Bilayer Graphene}

\renewcommand{\theequation}{S\arabic{equation}}
\renewcommand{\thefigure}{S\arabic{figure}}
\renewcommand{\thesection}{S\arabic{section}}
\renewcommand{\bibnumfmt}[1]{[S#1]}
\renewcommand{\citenumfont}[1]{S#1}
\newcommand{\absVal}[1]{\left|#1\right|}

\setcounter{figure}{0}

In TDBG, the mechanism of flat band formation is different from that of TBG, as noted in the main text. The nature of the low energy bands can be understood qualitatively by first considering the limit in which the inner two graphene layers, the layers which have a relative twist, are decoupled. As illustrated in Fig. S1(a), the low-energy states at the moir\'e Brillouin zone (MBZ) corners are then localized on individual layers from 1 to layer 4, and on the honeycomb sub-lattice in that layer which does not have a near-neighbor in the same bilayer.  The energies at the MBZ corners are $3\Delta_V/2$, $\Delta_V/2$, $-\Delta_V/2$ and $-3\Delta_V/2$ respectively, where $\Delta_V$ is the on-site energy difference between adjacent layers. While in TBG, low-energy states localize on both sub-lattices of each layer. The color-coded bars ($y$-label on the right) in Fig S1 give the layer-dependent probability distributions for the corresponding high-symmetry $k$-points in the MBZ marked by colored points in the band structure. When the inner layers are then coupled, level repulsion tends to push the energies of states with high weight in these layers to high energies.   The interplay between sublattice-coupling within layers and sublattice-dependent interlayer tunneling which yields the magic angle flat bands in bilayers does not take place because only one sublattice is available in each layer at low energies.  This leaves low energy bands that are dominantly localized in the outer two layers, as illustrated in Fig. S1(b).

At $n = n_s/2$, the screened potential energy difference between two adjacent graphene layers is $\Delta^\epsilon_V = \frac{4\pi e^2 nd}{\epsilon} = \frac{2\pi e^2 n_s d}{\epsilon} \approx \frac{145 \text{meV}}{\epsilon}$, where $d = 0.34$nm, and the superlattice density $n_s \approx 4\times 10^{12}$ cm$^{-2}$ for $\theta=1.33^\circ$. Using a relative permittivity of twisted bilayer graphene of $\epsilon \sim 5$ \cite{chung_transport} leads to a screened potential energy $\Delta^\epsilon_V \approx 30$meV.

\begin{figure*}[!h]
\centering
\includegraphics[width=\linewidth]{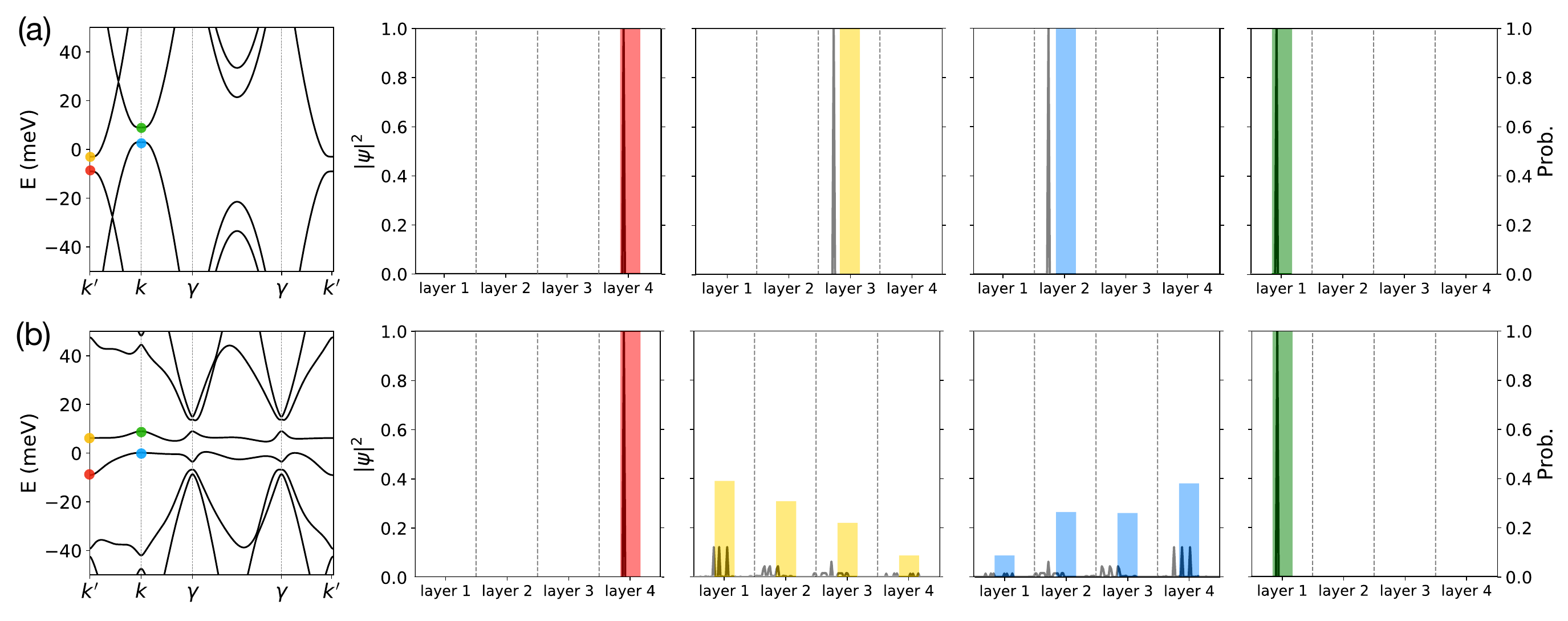}
\caption{\small{TDBG moir\'e bands neglecting (a) and including 
(b) coupling between the middle two layers which have a relative twist $\theta=1.01\degree{}$.  Both (a) and (b) were calculated at $\Delta_V=5$ meV, where $\Delta_V$ is the on-site energy difference between adjacent layers. Here the top two layers (1,2) have higher on-site energies than the bottom two layers (3,4). (a) Band structure and layer probabilities at specific high-symmetry points (colored dots) without coupling between graphene layer 2 and layer 3. (b) With coupling between graphene layer 2 and layer 3.}}
\label{fig S1}
\end{figure*}

\begin{figure}[H]
\centering
\includegraphics[scale=0.55]{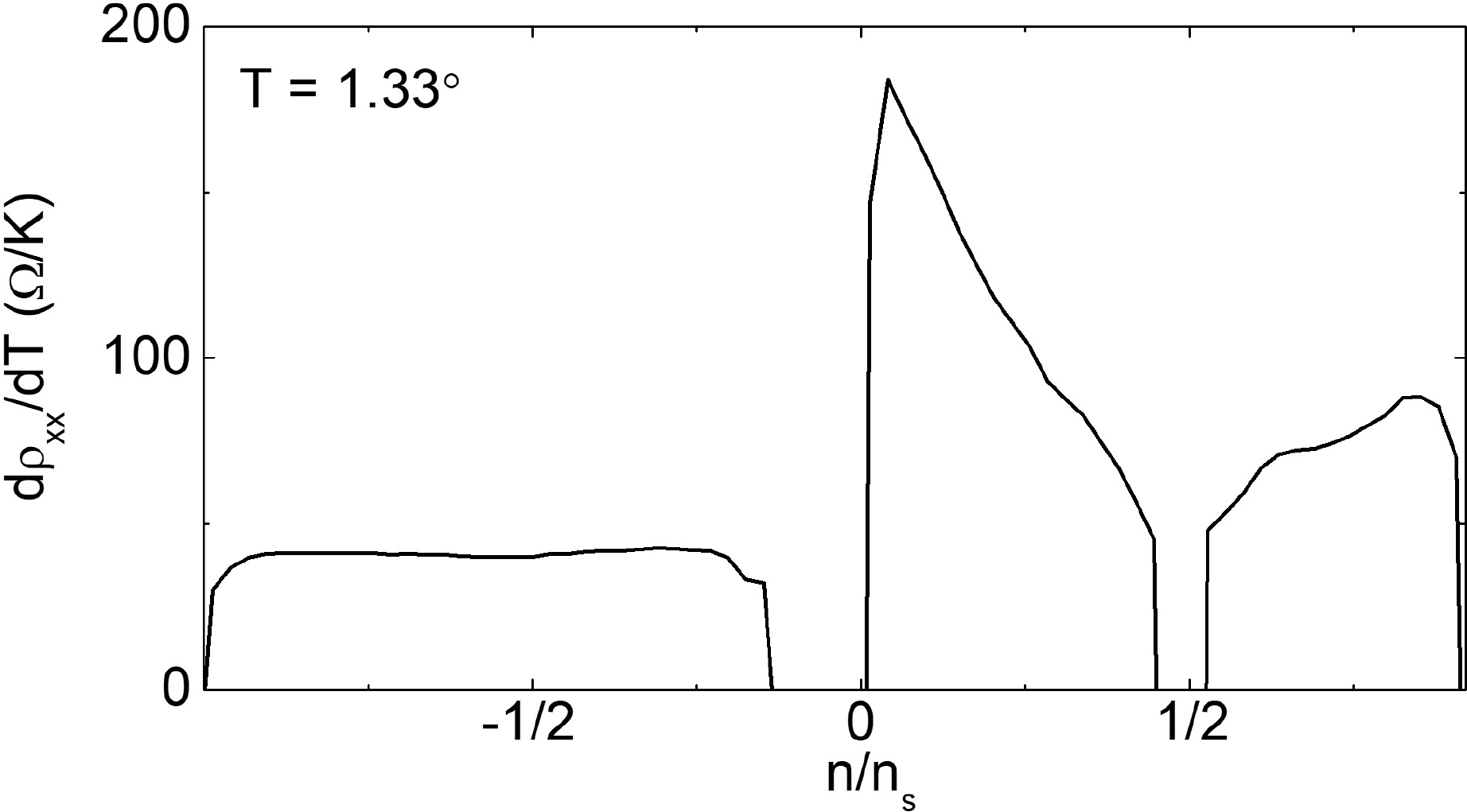}
\caption{\small{d$\rho_\mathrm{xx}$/d$T$ vs. $n$ in units of $n_\mathrm{s}$ for the TDBG sample with $\theta = 1.33\degree$, in the linear region of the temperature dependence ($T < \SI{50}{K}$).}}
\label{fig S2}
\end{figure}

\begin{figure}[H]
\centering
\includegraphics[scale=0.6]{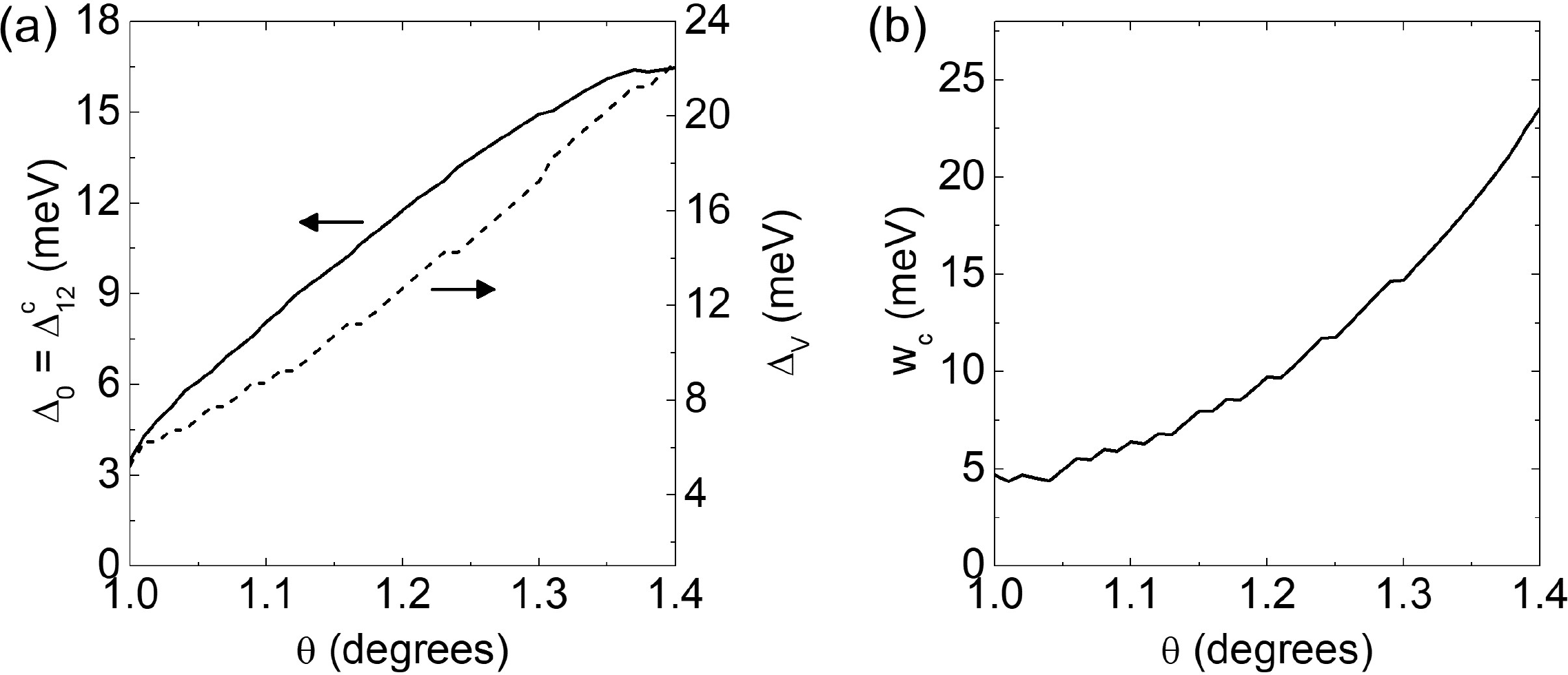}
\caption{\small{(a) Calculated band gaps at charge neutrality ($\Delta_0$) and between the first and second conduction bands ($\Delta_{12}^c$) (solid) and on-site energy difference ($\Delta_\mathrm{V}$) due to an applied electric field (dashed) in TDBG at the condition $\Delta_0=\Delta_{12}^c$, as a function of twist angle ($\theta$). (b) Calculated band width of the first conduction band ($w_c$) at $\Delta_0=\Delta_{12}^c$, as a function of $\theta$.}}
\label{fig S3}
\end{figure}

\begin{figure}[H]
\centering
\includegraphics[scale=0.65]{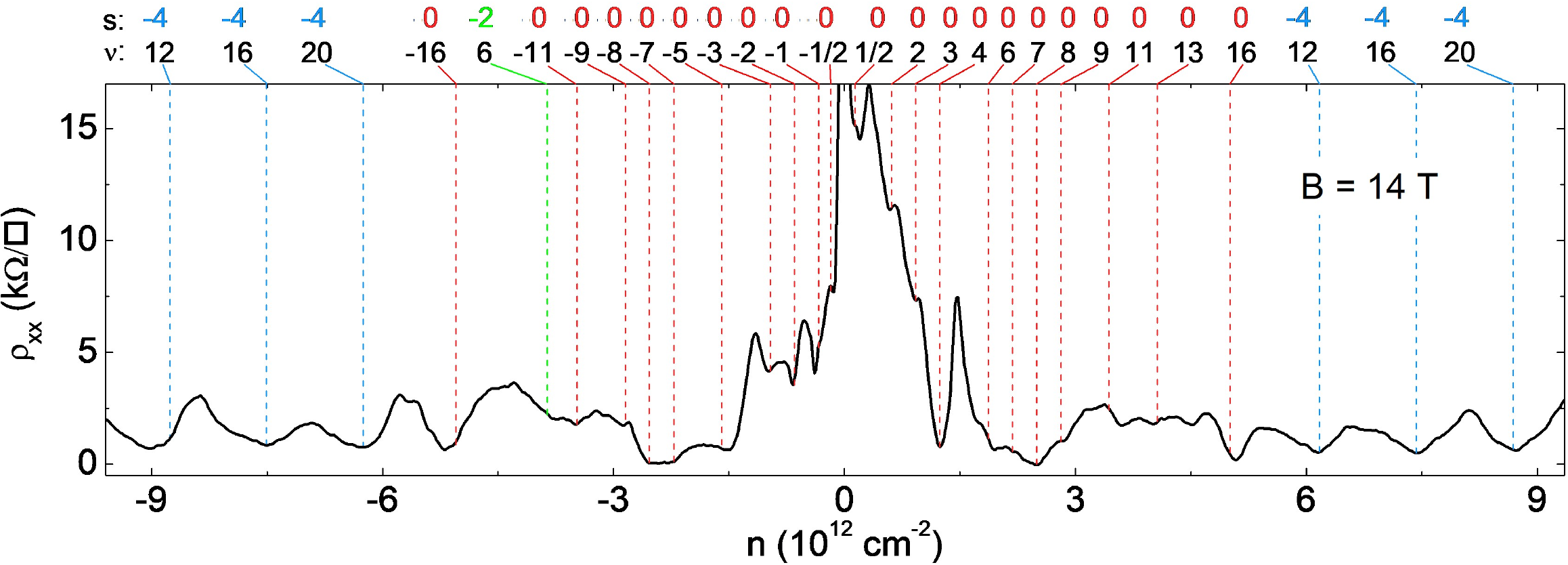}
\caption{\small{$\rho_\mathrm{xx}$ vs. $n$ measured at $B = \SI{14}{T}$ in the TDBG sample with $\theta = 1.10\degree$, and $T = \SI{0.3}{K}$ [Line-cut of Fig 4(a)]. $\rho_\mathrm{xx}$ minima are labeled by dashed lines according to originating moir\'e band ($s$) and Landau level filling factor ($\nu$). Fractional and single integer $\nu$ values in the fan corresponding to $s = 0$ indicate very low levels of disorder in the sample.}}
\label{fig S4}
\end{figure}


\end{document}